\documentclass[journal]{vgtc}                % final (journal style)

\usepackage{mathptmx}
\usepackage{graphicx}
\usepackage{times}
\usepackage[table,xcdraw]{xcolor}
\usepackage{subfig}
\usepackage{amsmath}
\usepackage{amsthm}
\usepackage{units}
\usepackage{amssymb}

\usepackage[ruled]{algorithm2e} % For algorithms

\usepackage{amsmath}
\usepackage{amssymb}
\usepackage{amsthm}
\usepackage{newlfont} % for Box
\usepackage{floatflt}
\usepackage{wrapfig}
\usepackage{fixltx2e}
\usepackage{subfig} % for subfloat
\usepackage{multirow}
\usepackage{booktabs}
\usepackage{algorithmic}
\usepackage[export]{adjustbox}
\usepackage{mathtools, cuted}
\usepackage{CJKutf8} % Chinese

\usepackage[ruled]{algorithm2e} % For algorithms

\usepackage[bookmarks,backref=true,linkcolor=black]{hyperref} %,colorlinks

\usepackage{cleveref}

\onlineid{1020}

\vgtccategory{Research}

\vgtcinsertpkg

\newcommand{\nothing}[1]{{}}
\newcommand{\shortcite}[1]{\cite{#1}}

\title{Gaze-Contingent Retinal Speckle Suppression \\ for Perceptually-Matched Foveated Holographic Displays}

\author{Praneeth Chakravarthula, Zhan Zhang, Okan Tursun, Piotr Didyk, Qi Sun, and Henry Fuchs}
\authorfooter{
\item
 Praneeth Chakravarthula is with University of North Carolina at Chapel Hill. E-mail: cpk@cs.unc.edu.
\item
 Zhan Zhang is with University of Science and Technology of China. E-mail: whirlwind@mail.ustc.edu.cn
\item
 Okan Tursun is with Università della Svizzera italiana. E-mail: tursuo@usi.ch
\item
 Piotr Didyk is with Università della Svizzera italiana. E-mail: piotr.didyk@usi.ch
\item
 Qi Sun is with New York University. E-mail: qisun@nyu.edu
\item
 Henry Fuchs is with University of North Carolina at Chapel Hill. E-mail: fuchs@cs.unc.edu.
}

\shortauthortitle{Chakravarthula \MakeLowercase{\textit{et al.}}: Gaze-Contingent Retinal Speckle Suppression for Perceptually-Matched Foveated Holographic Displays}

\abstract{Computer-generated holographic (CGH) displays show great potential and are emerging as the next-generation displays for augmented and virtual reality, and automotive heads-up displays. One of the critical problems harming the wide adoption of such displays is the presence of speckle noise inherent to holography, that compromises its quality by introducing perceptible artifacts. Although speckle noise suppression has been an active research area, the previous works have not considered the perceptual characteristics of the Human Visual System (HVS), which receives the final displayed imagery. 
However, it is well studied that the sensitivity of the HVS is not uniform across the visual field, which has led to gaze-contingent rendering schemes for maximizing the perceptual quality in various computer-generated imagery. 
Inspired by this, we present the first method that reduces the ``perceived speckle noise'' by integrating foveal and peripheral vision characteristics of the HVS, along with the retinal point spread function, into the phase hologram computation.
Specifically, we introduce the anatomical and statistical retinal receptor distribution into our computational hologram optimization, which places a higher priority on reducing the perceived foveal speckle noise while being adaptable to any individual's optical aberration on the retina. Our method demonstrates superior perceptual quality on our emulated holographic display. Our evaluations with objective measurements and subjective studies demonstrate a significant reduction of the human perceived noise.} % end of abstract


\CCScatlist{ % not used in journal version
 \CCScat{K.6.1}{Management of Computing and Information Systems}%
{Project and People Management}{Life Cycle};
 \CCScat{K.7.m}{The Computing Profession}{Miscellaneous}{Ethics}
}

  \teaser{
 \centering
 \includegraphics[width=16cm]{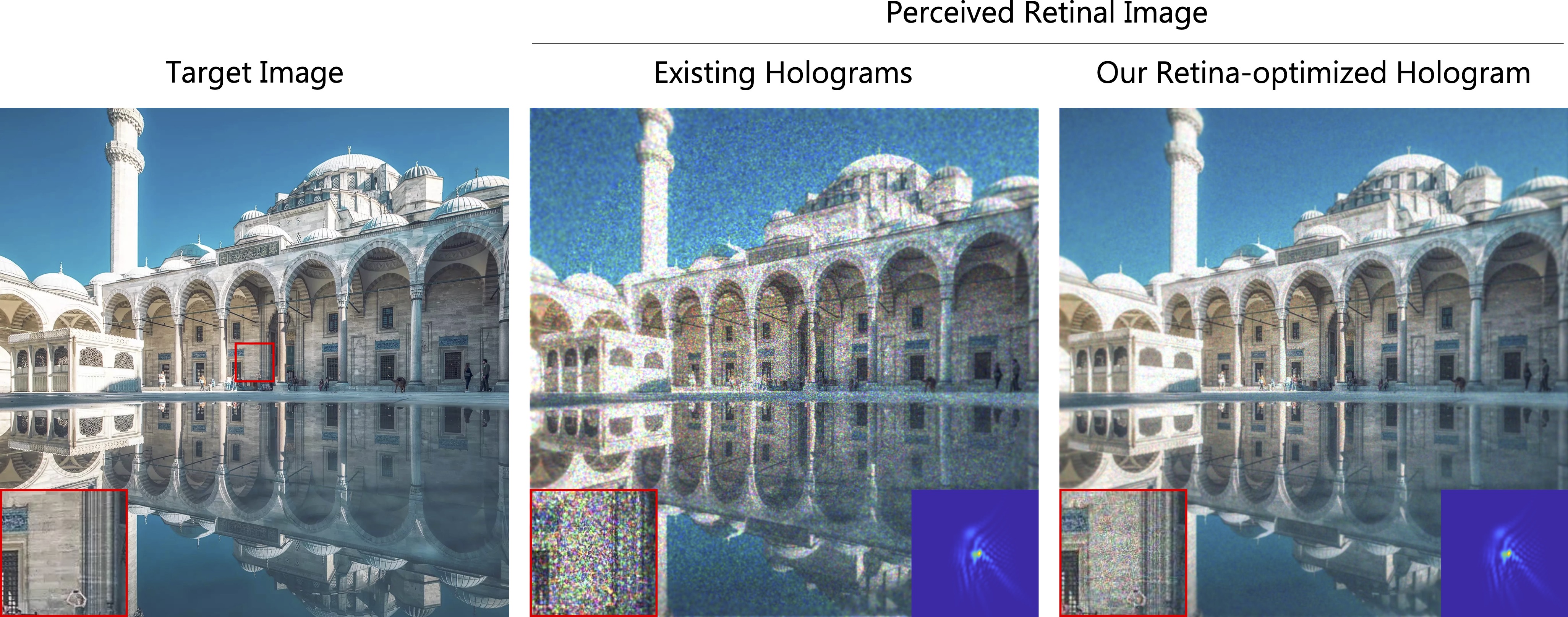}
  \caption{\emph{Visual comparison of our gaze-contingent retinal phase optimization}. As the quality reference, the first image shows the target spatial image to be reconstructed by holograms. The second shows the perceived retinal image with existing holographic phase retrieval methods~\cite{chakravarthula2019wirtinger,peng2020neural}. The third shows our gaze-contingent retinal reconstruction assuming user's gaze in the middle of the screen.
  The inset figures show zoom-in foveal views and corresponding human eye point-spread-function used in the simulated reconstruction.
  }
  }

\begin{document}

\firstsection{Introduction}

\maketitle

Computer-generated holographic (CGH) displays offer several advantages over conventional projectors, such as immensely reduced optical complexity and on-demand light steering, and they are a promising technology for future augmented and virtual reality (AR/VR) and automotive heads-up displays (HUDs). One of the multitude of advantages that holographic displays offer is their ability to delegate several of the optical complexities from hardware to computation. For example, conventional near-eye AR and VR displays are often bulky due to reflective and refractive optics needed for projecting images to comfortable distances for viewing while simultaneously compensating for severe optical aberrations~\cite{chakravarthula2018focusar, xia2019towards}. On the other hand, a holographic display is capable of dynamically moving the image projection distance and also compensating for severe optical distortions, all in the hologram computation process~\cite{maimone2017holographic}. 

However, the image quality of existing holographic displays is modest compared to conventional displays. Apart from potential device imperfections, the use of either phase-only or amplitude-only spatial light modulators (SLM) also results in degraded image quality, which because SLMs cannot perform the complex wave modulation required for holographic image formation. Owing to its improved diffraction efficiency, a phase-only SLM is typically used for holography; however, it requires devising high-quality holographic phase retrieval algorithms. 

Researchers have proposed several phase retrieval algorithms in the past that can be used for computing phase-only holograms for a holographic display. These include one-step hologram computation such as the double phase encoding approach~\cite{hsueh1978computer}, iterative heuristic phase optimization such as the Gerchberg-Saxton (GS)~\cite{gerchberg1972practical} and Fienup~\cite{fienup1982phase} methods and error diffusion algorithms~\cite{barnard1988optimal}. While these algorithms produce reasonable reconstructions, they are still noisy. More advanced recent methods utilize optimization and machine learning~\cite{chakravarthula2019wirtinger,chakravarthula2020computing,peng2020neural,Shi2021:Nature,eybposh2020deepcgh} for directly optimizing the holographic phase patterns for overall image quality. For specific applications, such as augmented and virtual reality, it is neither acceptable to display noisy images nor necessary to compute extremely high-resolution noise-free imagery. What we need are holographic projections that are perceptually noise-free and high resolution.

Looking at the human visual system (HVS), the optics of the eye is imperfect causing non-sharp point spread functions (PSF) on the retina. An imperfect PSF causes self interference of the diffracted field resulting in speckle noise. However, photoreceptors on the retina are distributed non-uniformly, with the highest density in the fovea and progressively lower densities with increasing eccentricity in the periphery. This results in different levels of visual sensitivities in the foveal and peripheral vision. Foveal vision is characterized by a high visual acuity, whereas a loss of spatial resolution of perceived imagery occurs with increasing eccentricity in the peripheral vision. As a result, the perception of higher spatial frequencies in the images, such as in speckle noise, is suppressed in the peripheral visual field. Although foveal vision has superior acuity, it spans only about a modest 6 degrees of field of view. This non-uniformity of photoreceptors facilitates additional degrees of freedom in terms of phase hologram computation, where noise can be redistributed to imperceptible regions in the periphery while optimizing for the image quality in the fovea. 

In this paper, we address this under-investigated topic of computer generated holography, and we introduce a systematic anatomical model of retinal receptor density to the optimization-based SLM phase retrieval, aiming at reducing the {\it perceived} noise on the retina. We use this anatomical model to optimize for phase holograms that are of high perceptual quality. 
Our scalable algorithm also supports both generalization and individualization of retinal optical aberrations i.e., PSFs. 
Furthermore, we analyze the benefits of such perceptually-aware hologram synthesis with both objective image-space metrics and subjective psychophysical studies. We show that such perceptually-aware holographic projections are comparable in quality to holographic projections optimized over the entire image. 
We envision this to be an essential step towards designing algorithms to produce perceptually superior quality holograms for future displays.

In specific, we make the following contributions:\vspace{-0.5em}
\begin{itemize}
    \item A perceptually high-quality holographic phase retrieval method with perceptually suppressed speckle noise.\vspace{-0.5em}
    \item A gaze-contingent algorithm that introduces analytical retinal receptor models to holographic phase retrieval optimization.\vspace{-0.5em}
    \item An optimization method that compensates for viewer's retinal aberrations via decoupling PSF in the computation procedure.\vspace{-0.5em}
    \item A series of perceptual studies on an emulated holographic display\footnote{Imperfections in holographic display setups, such as aberrations in display optics or non-linearities in phase modulating SLM, causes noise and degradation in image quality irrespective of the choice of phase retrieval algorithm. To decouple the effects of retinal-optimized holograms and any real-world errors, we conduct our user evaluations on an emulated holographic display. Using an emulated holographic display allowed us to better understand the perceptual effects of gaze-contingent retinal speckle suppression in isolation, eliminating any other sources of error.}.
\end{itemize}

\section{Related Work}
In this section, we review relevant holographic display technologies and computer-generated holography (CGH) algorithms.  

\subsection{Holographic Displays}
Holographic displays are gaining increasing interest as the ultimate display technology, promising compact 3D displays with per-pixel focus control and aberration correction, which are essential features especially for future immersive near-eye displays. Most holographic display designs rely on a single phase-only spatial light modulator (SLM)~\cite{chen2015improved}, although configurations using two phase SLMs~\cite{levin2016passive,chakravarthula2019wirtinger,choi2021optimizing}, or a combination of both amplitude and phase SLMs~\cite{shi2017near} are also explored. While typically holographic displays have been large and bulky benchtop setups, recent works have shown path to miniaturization of such displays via holographic optics HOEs~\cite{maimone2017holographic, li2016holographic, jang2019holographic} or waveguides~\cite{yeom20153d}. Two critical limitations of existing holographic near-eye displays are the tiny eyebox and poor image quality. Several recent works have achieved an increased eyebox size by using eyetracking~\cite{jang2017retinal, jang2019holographic,xia2020towards}. While the display design plays a major role in the final image quality, existing designs only differ in the specific implementation details. In contrast, the algorithms used for generating the holograms, which restrict the achievable image quality, remain the same or similar across the displays. However, recent optimization and deep learning approaches have shown significant algorithmic improvements in image quality~\cite{chakravarthula2019wirtinger,chakravarthula2020learned,Shi2021:Nature,peng2020neural}.

\begin{figure*}[thb]
\centering
    \includegraphics[width=0.8\linewidth]{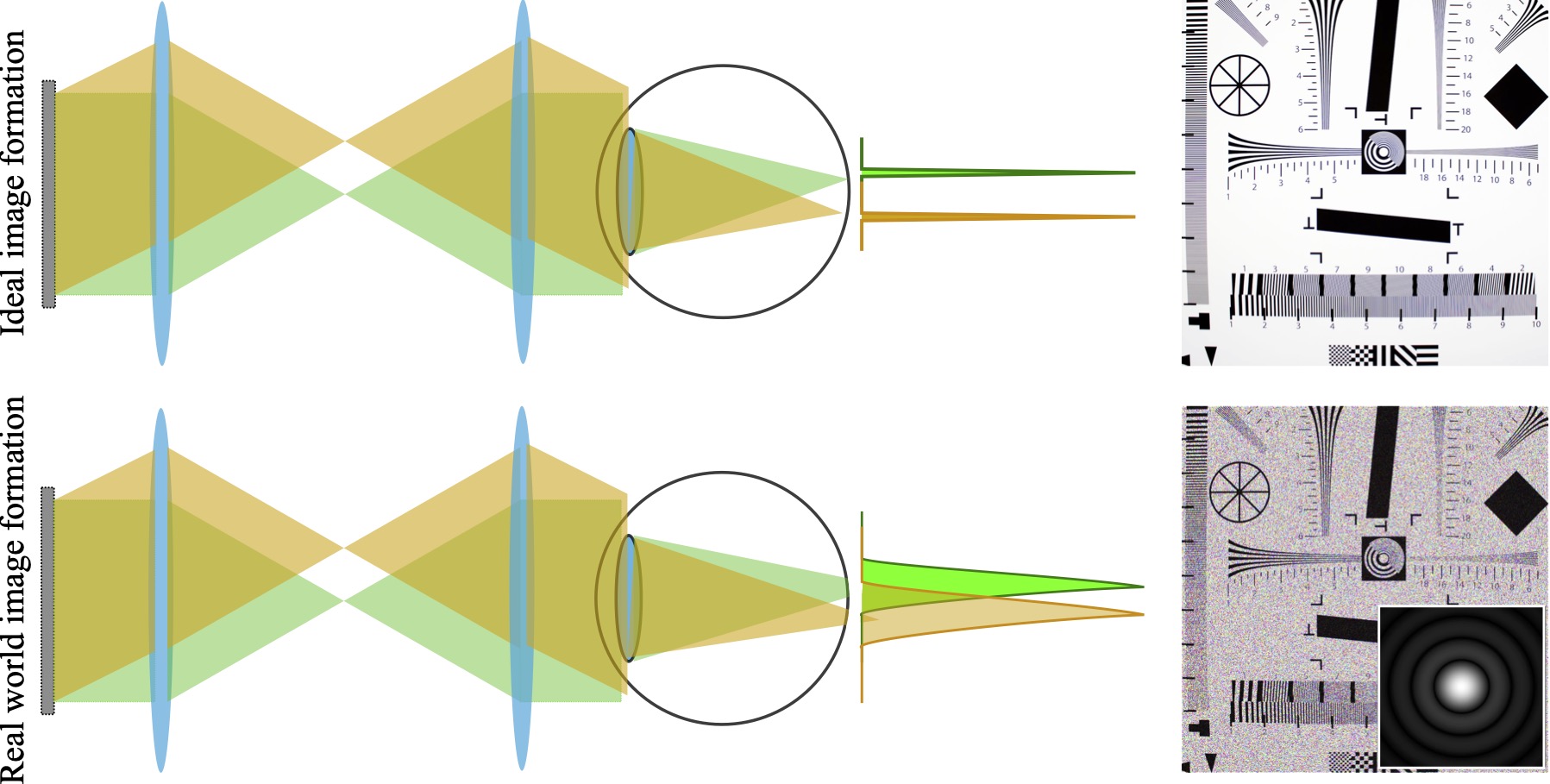}
    \caption{Speckle noise due to PSF on the image plane: [Top] Simulated holographic reconstructions as perceived by a human observer when there is no overlap in the image plane (complex amplitude) pixels. [Bottom] Simulated holographic reconstructions as perceived by a human observer when there is an overlap in the image plane (complex amplitude) pixels. Overlapping (complex amplitude) image pixels with random phase result in interference and causes speckle noise. [Inset] Visualization of Airy disk for a circular aperture.}
    \label{fig:speckle}
\end{figure*}

\subsection{Limitations in Holographic Display Setups}

In this section, we briefly review holographic display configurations and the most important limitations that restrict image quality in holographic displays. Note that owing to these several factors deteriorating image quality in holographic displays, we are unable to separate the effects of speckle noise due to retinal PSF on a real holographic display. Therefore, our user evaluations are conducted on an emulated holographic display. 

\paragraph{SLM technology}
Phase-only SLMs are often preferred for computer generated holography due to their higher diffraction efficiency. Achieving accurate, complex modulation of both phase and amplitude with a single SLM device still remains an open problem~\cite{reichelt2012full}. However, these phase-only SLMs require a trillion sub-wavelength sized pixels to display holograms comparable to conventional displays. Unfortunately, existing SLMs only have resolutions ranging up to $3840 \times 2160$ (4K UHD) with pixel pitches limited to approximately $\unit[4]{\mu m}$ and fill factors less than $95\%$. While decreasing pixel pitches result in increasing maximum diffraction angle and thus field of view, it also causes a significant drop in the image quality. The space for electronics between the active pixel areas further leads to zero-order undiffracted light which often causes severe artifacts in the holographic images.

\paragraph{Phase wrapping and quantization errors}
Most existing SLMs can only present quantized phase modulation with a limited bit-depth. The CGH phase patterns are typically wrapped to lie within $[0, 2\pi]$, which are further quantized to the bit-depth of the SLM. The phase errors caused due to quantization and wrapping operations significantly deteriorate the quality of holographic images~\cite{dallas1972phase}. This quantized digital modulation of the SLM pixels correspond to a phase modulation of the incident light via a calibrated lookup table (LUT), which determines the phase modulation accuracy and diffraction efficiency of the SLM. Any errors in the LUTs cause non-linear phase modulation which further degrade the display quality.

\paragraph{Coherent Laser Speckle}
Illumination sources such as single-mode lasers have a large coherence length and produce coherent speckle noise. Many works aim at reducing speckle by using rotating diffusers~\cite{bianco2016quasi}, modulating or quickly repositioning the laser beam~\cite{kang2008effective} or superposition of multiple reconstructions~\cite{golan2009speckle}. While such coherence speckle noise can also be mitigated by using partially-coherent light sources, they result in blur or loss of depth perception~\cite{dainty1977statistics}. For a comprehensive discussion on many techniques to reduce speckle noise in CGH, we refer the reader to Bianco et al.~\cite{bianco2018strategies}. However, note that effective holographic noise suppression still remains an open problem.

\subsection{Iterative Algorithms for CGH} 
Holography for displays rely on diffraction and interference of light for generating imagery. Using a phase-only SLMs requires computing phase-only holograms that are capable of projecting imagery that can closely mimic the target image. The phase retrieval problem is generally non-convex and ill-posed. Early methods to phase retrieval include error reduction using iterative optimization~\cite{lesem1969kinoform,gerchberg1972practical} and hybrid input-output (HIO) methods~\cite{fienup1982phase, bauschke2003hybrid}. First-order non-linear optimization~\cite{gonsalves1976phase,lane1991phase,fienup1993phase}, alternative direction methods for phase retrieval~\cite{wen2012alternating,marchesini2016alternating}, non-convex optimization~\cite{zhang20173d}, and methods overcoming the non-convex nature of the phase retrieval problem by lifting; i.e., relaxation, to a semidefinite~\cite{candes2013phaselift} or linear program~\cite{goldstein2018phasemax,bahmaniPhaseMax} are also explored. Recent optimization methods for holographic phase retrieval utilize Wirtinger gradients and stochastic gradient descent methods to compute high-quality holograms with flexible loss functions~\cite{chakravarthula2019wirtinger, chakravarthula2020computing,peng2020neural}. Deep learning based methods are also gaining interest among researchers for their ability to predict high-quality phase-only holograms~\cite{peng2020neural, chakravarthula2020computing,Shi2021:Nature,eybposh2020deepcgh}. We refer the reader to Barbastathis et. al~\shortcite{barbastathis2019use} for an overview of learned phase retrieval methods. 

All of these methods have in common that they assume a perfect image formation model and ignore deviations from the perfect forward model in the human eye. The proposed method addresses this very important but under-investigated aspect of holographic displays.

\section{Computer Generated Holography}
\label{sec:cgh}
\begin{figure*}[t]
    \centering
    \includegraphics[width=1.02\linewidth]{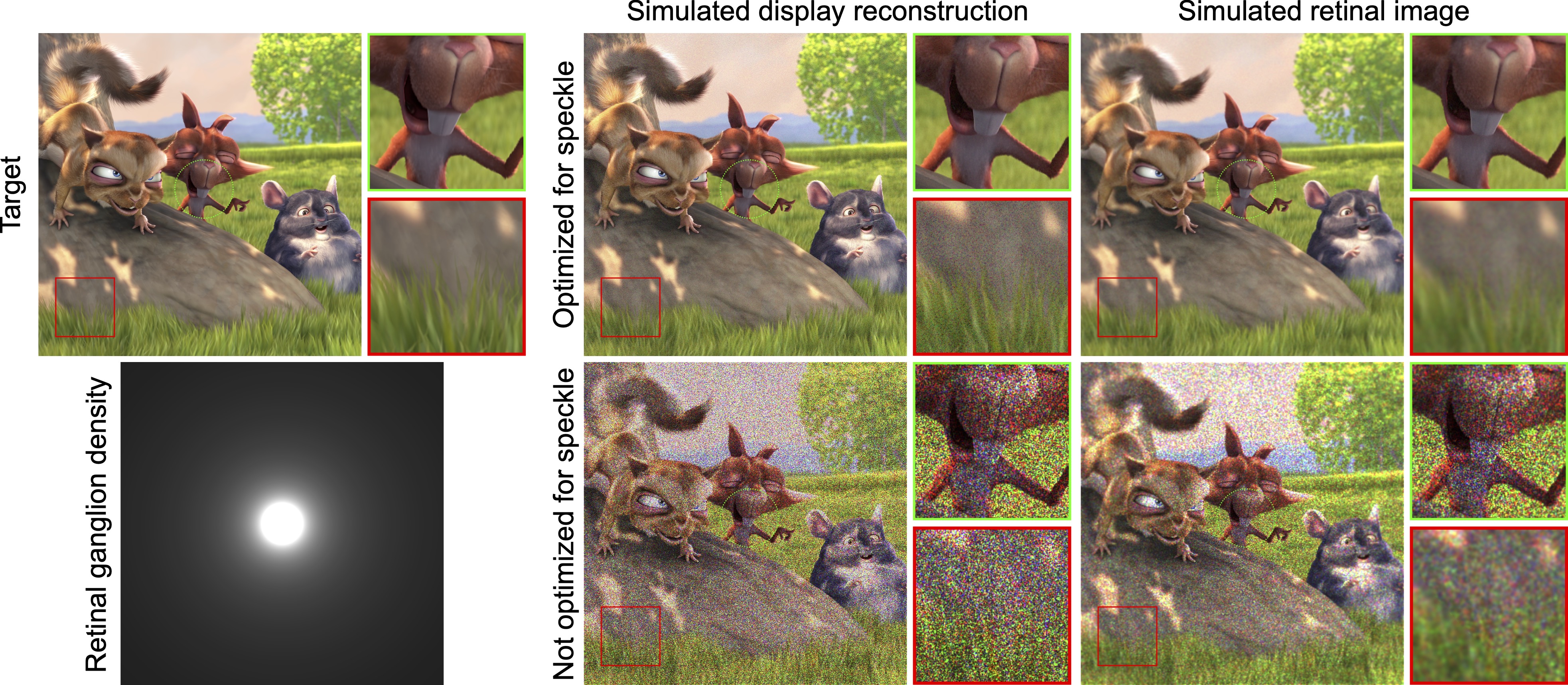}
    \caption{Comparison of simulated reconstructions and retinal images from two different optimization methods with and without using retinal sampling model and PSF. Foveation with a retinal sampling model and PSF optimization shifts the noise towards peripheral visual field, where it is less visible.}
    \label{fig:noiseVis}
\end{figure*}    
In this section, we introduce computer-generated holography and holographic image formation in displays, and briefly discuss both near-field Fresnel holograms and far-field Fourier holograms. In later sections, we discuss in detail our approach to perceptually-aware hologram synthesis.

A holographic display creates imagery from the interference of diffracted light, as opposed to conventional displays. While the conventional LCD-based displays are limited by diffraction, holographic displays take advantage of diffraction to form images. A holographic display typically utilizes a spatial light modulator (SLM) which modulates either the amplitude or phase of the incoming light. The modulated wavefront then propagates to a certain distance, where the interference pattern results in the intended image. Computer-generated holography aims to simulate this real-world interference and calculate the corresponding wavefront modulation required on the SLM plane. 

As discussed earlier, a fundamental limitation of existing holographic displays is the unavailability of efficient complex wave modulating SLMs. In a typical holographic display, a phase-only SLM displaying a phase pattern $\Phi$ is illuminated by a wavefield generated by a coherent (laser) source $U_{s}$, to generate a complex-valued wave field $U_H = U_{s} e^{j\Phi}$ on the hologram plane. This way, the phase of the input illumination field is modulated by the displayed SLM phase $\Phi$. The modulated field propagates in free space to the target image plane, where the intensity of this propagated wave field $U_I$ is observed (i.e., perceived by a user or measured by a camera).

The free space propagation of modulated wave field can be described by the scalar diffraction integral:
\begin{equation}
	U_I(x, y) = \frac{1}{j\lambda} \int_{-\infty}^{\infty} \int_{-\infty}^{\infty} U_{H}(\zeta, \eta) \frac{\mathbf{exp}(jk\rho)}{\rho} d\zeta d\eta,
	\label{eq:diffractionIntegral}
\end{equation}
where $(\zeta, \eta)$ are the coordinates on the hologram plane, $(x, y)$ are the coordinates on the image plane, $k=\frac{2\pi}{\lambda}$ is the wave number and $\rho = \sqrt{(\zeta - x)^2 +(\eta - y)^2 + d^2}$ is the Euclidean distance between the points on the hologram and image planes, respectively. Note that a phase-only hologram modulates only the phase of light, and hence has a constant amplitude across the hologram plane which is nothing but the illumination source intensity. Although this illumination intensity is different for different hardware configurations, it is generally assumed to be unity for the purpose of phase retrieval. 

While the scalar diffraction integral as described above in \Cref{eq:diffractionIntegral} is perhaps the most accurate description of the propagating wave field, it is computationally expensive. However, several simplifying assumptions can be made based on the propagation distance of the wave field, to make the above integral computationally tractable. For example, the diffraction integral can be interpreted as plane waves traveling in different directions from the hologram (SLM) plane, reformulating it into the angular spectrum propagation representation:
\begin{equation}
	U_I(x, y) = \int \int \mathcal{F} \big( U_{H}(\zeta, \eta) \big) \mathcal{H}(f_{\zeta}, f_{\eta}) e^{j2\pi (f_{\zeta}\zeta + f_{\eta}\eta)} df_{\zeta} df_{\eta}
\end{equation}

\begin{equation}
\mathcal{H}(f_{\zeta}, f_{\eta}; z) = 
\begin{cases} 
\exp{ \Big [ j2\pi\frac{z}{\lambda} \sqrt{1-(\lambda f_{\zeta})^2 - (\lambda f_{\eta})^2} \Big ] } &\hspace{-3mm}, \sqrt{f_{\zeta}^2 + f_{\eta}^2} < \frac{1}{\lambda} \\
0 &\hspace{-3mm}, \text{otherwise,}
\end{cases}
\end{equation}
where $\mathcal{F}$ denotes the Fourier transform operator, $\lambda$ is the wavelength, $f_{\zeta}, f_{\eta}$ are corresponding spatial frequencies, $z$ is the distance between the hologram (SLM) plane and the image plane. Similarly, Fresnel and Fraunhofer or Fourier approximations can be invoked to describe near field and far field wave propagation, respectively~\cite{goodman2005introduction}. For the ease of description and calculation, we use far field Fourier holography in this work, where the wave propagation is described by a Fourier transformation:
\begin{equation}
    U_I(x, y) = \mathcal{F} \big[ U_{s}(\zeta, \eta) e^{j\Phi(\zeta, \eta)} \big]
\end{equation}
Given that we only observe the intensity of the wave field on the image plane, i.e. $|U_I|^2$, the problem of synthesizing phase holograms can be thought of as computing the SLM phase pattern $\Phi$ such that the observed intensity $|U_I|^2$ matches the target image intensity. In other words, phase hologram computation is equivalent to solving the following optimization problem:
\begin{equation}
\begin{aligned}
\Phi_\text{opt} = \hspace{1.5mm}
& \underset{\Phi}{\text{minimize}}
& & \textstyle \mathcal{L}( |U_I(\Phi)|^2, I )
\label{eq:basic-optim-problem}
\end{aligned}.
\end{equation}
where $\mathcal{L}$ is a penalty function that measures the error between the reconstructed image intensity and the target image intensity. As can be seen, solving the above phase retrieval problem is nothing but an iterative refinement of the SLM phase pattern until the error between the reconstructed holographic image and the target image is minimized. Traditional heuristic \emph{ping-pong style} iterative phase retrieval algorithms such as Gerchberg-Saxton or Fienup methods can be thought of as solving the above minimization problem to compute the required SLM phase pattern~\cite{bauschke2002phase}. Recent methods such as Wirtinger Holography~\cite{chakravarthula2019wirtinger} solves the above nonconvex optimization problem using first-order gradient descent methods, resulting in superior holographic reconstruction quality. However, the observed images from a real hardware holographic display are generally distorted with visible speckle noise.

\vspace{2mm}

\noindent
\textbf{Speckle noise in holographic images} from a real holographic display is caused from a variety of sources, few of them being nonuniformities on the SLM pixel surface, diffuse surfaces in the optical path and nonuniform illumination source. However, the major source of speckle noise in holographic displays is due to the aperture size of devices in the optical system, which causes a point spread of function (PSF) on the retina. This PSF extends over an area, generally called the Airy disk. Note that the Airy disk is the diffraction pattern resulting from a circular aperture. For a far-field hologram (which is far from the aperture), the angle at which the first minimum of the Airy disk occurs can be approximated by $1.22 \lambda/ D$, where $\lambda$ is the wavelength of light, and $D$ is the aperture size of the display system. Furthermore, this diffraction pattern of the aperture is additionally characterized by the sensitivity of the eye.

If the diameter of this Airy disk, referred to as PSF hereby, is larger than the interval of pixels on the image plane, then the neighboring pixels overlap, causing interference. If the phase on the image plane is uniform, this would only result in a spatially blurred holographic image projection. However, if the phase distribution on the image plane is random, the addition of the complex amplitude of neighboring pixels (interference) on the image plane causes speckle noise, as shown in \Cref{fig:speckle}. We simulate the effect of speckle from the interference of overlapping pixels by convolving the image plane complex wavefield with a PSF function. As can be observed from \Cref{fig:speckle}, while non-overlapping image pixels produce noise-free holographic image reconstructions, any overlap will quickly result in perceivable speckle noise if the phase on the image plane is random. In Section~\ref{sec:method}, we discuss our approach to reduce perceived speckle by incorporating the PSF into the retinal image formation model for generating the phase holograms.

\section{Retinal Sampling Model and the Point Spread Function of the Human Eye}

\label{sec:hvs_sampling}

Optical and neural processes involved in the HVS for the perception of a visual stimulus has been the subject of many studies in the past in order to steer image reconstruction algorithms towards improving the \emph{perceived quality}, as opposed to noise due to pixel-wise differences. The former has a direct impact on the user experience while the latter has the possibility of performing costly computational optimizations without any perceptible quality improvements for a human observer. Such studies introduced different models of the HVS which explain the underlying mechanisms behind the perception of fundamental characteristics of visual stimuli such as color, motion and depth, as well as higher level perceptual phenomena such as visual masking, attention and visual processes involved in object recognition.

Visual perception starts with the optical image formation on the human retina by the eye's crystalline lens. This retinal image is later converted to a neural representation by the array of photoreceptor cells on the retina, a process which shares similar characteristics with image acquisition performed by the image sensors found in traditional cameras. Similar to cameras, the level of spatial detail that can be resolved by the human eye is limited and is mainly driven by optical aberrations of the eye's lens, and Nyquist sampling rate defined by the density of photoreceptor cells on the retina. However, the most striking difference between a camera and a human eye is that the density of the photoreceptor cells is not uniform across the visual field, which results in different sampling rates for the visual signal in the central vision (fovea), where the photoreceptor cells are more densely packed and in the periphery, where the density and sampling rate is low. As a consequence, the human eye can resolve a higher level of spatial details in the foveal region compared to the periphery.

Most perceptual signal processing applications define a bandwidth for the image signal representation by taking into account the capability of the human retina for resolving spatial signals. Earlier studies confirmed the difference in spatial resolution limits of the foveal and peripheral vision using psychovisual experiments~\cite{hering1899uber,levi1986sampling}. This observation sparked interest in studying the topology of the human retina and the distribution of photoreceptor cells such as cones and midget retinal ganglion cells (mRGC) which define the neural limits of spatial resolution. Initial sampling of the visual signal is performed by cone cells on human retina and the signal is later transmitted from cones to the early stages of the visual cortex by mRGC. Studies show that the mRGC to cones ratio (mRGC/cones) is approximately two in the fovea, where the spatial resolution limit is mainly determined by optical aberrations and the density of cones~\cite{kolb1991midget}. However, as the retinal eccentricity increases this ratio rapidly drops below one, indicating multiple cone connections to each mRGC~\cite{curcio1990topography}. As a result, the density of mRGC becomes critical for the spatial resolution power of the human retina~\cite{rossi2010relationship}.

Motivated by the close relation between the mRGC density and the neural sampling limits, Watson~\cite{watson2014formula} derived the formula for computing the density of mRGC as a function of retinal eccentricity as:  

\begin{equation}
    \label{eq:density}
    \rho(r, m) = 2\rho_{cone}(1+\frac{r}{41.03})^{-1}[a_m(1+\frac{r}{r_{2,m}})^{-2}+(1-a_m)e^{-r/r_{e,m}}],
\end{equation}
where $\rho(r,m)$ is the cone density at eccentricity $r$ degree
along meridian $m$, $\rho_{cone}=14,804.6\,{deg}^{-2}$ is the density of cone cell at fovea and $m\in\{1,2,3,4\}$ is index of four meridians of the visual field, and $a_m,r_{2,m},r_{e,m}$ are fitting constants listed in~\cite{watson2014formula}. In this work, we utilize the above model for measuring the visibility of the speckle noise across the visual field. Specifically, we aim to generate phase holograms such that the foveal region in the visual field achieves the highest fidelity reconstruction while the speckle noise is moved towards periphery where the neural sampling rates are lower. Consequently, the amount of perceived speckle noise in the visual field is reduced. We discuss our approach to gaze-contingent speckle reduction in Section~\ref{sec:method}.

\begin{figure}
\centering
    \includegraphics[width=0.95\columnwidth]{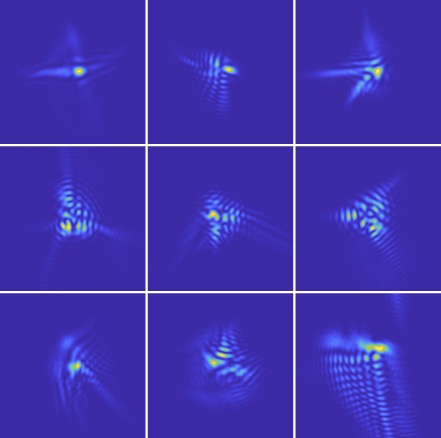}
    \caption{Sampled real human retina PSFs tested in our experiments. Retinal PSFs take different shapes as caused by the aberrations due to non-ideal optics of the eye \cite{thibos2002, Watson2015PSF}.}
    \label{fig:retina_psf}
\end{figure}

As mentioned in Section~\ref{sec:cgh}, the speckle noise perceived by the observer is influenced by the PSF of the human eye's optics, which has a variability between individuals. Thibos et al. \cite{thibos2002} investigated the statistical variability of the human PSF among healthy individuals and the amount of high-order aberrations in comparison with defocus and astigmatism. To this end, they measured the PSF of left and right eyes from 100 individuals using Shack-Hartmann wavefront sensor \cite{shack1971}. The collected wavefront aberration data is expressed as a weighted sum of Zernike polynomials \cite{thibos2000}. Based on this dataset, Watson later introduced a software to compute human optical PSFs from wavefront aberrations, including the effects of varying pupil diameter, wavelength of the light and the amount of aberrations such as defocus and astigmatism \cite{Watson2015PSF}.

Modeling PSF is critical for the perceived quality when optimizing the SLM phase pattern as shown in \Cref{fig:noiseVis}. While this requires user specific PSF measurements as shown in \Cref{fig:retina_psf}, it is possible to approximate an average PSF for the general population by using a Gaussian kernel~\cite{coppens2004new} and we show that the approximation is a practical solution when the actual PSF of the viewer is unknown. The optimization performed using the Gaussian approximation provides a better noise suppression performance than the existing holograms as we show in our evaluations even for the actual PSF measurements from Thibos et al. and Watson's dataset \cite{thibos2002, Watson2015PSF}. Therefore, the optimization performed using the approximated PSF already provides a noticeable visual quality improvement for a typical viewer. Nevertheless, our approach is not limited to a particular set of PSF and it also provides the flexibility of using a custom PSF during the optimization when the actual PSF of the viewer is known. This is especially useful if the viewer's PSF deviates significantly from the general population and optimizing for the individual PSF provides a more accurate representation of the high-order aberrations. Therefore, it is possible to obtain an improved speckle noise reduction on the per-viewer basis using the actual PSF measurements when they are available.

\section{Method}
\label{sec:method}

\begin{figure*}[t!]
    \centering
    \includegraphics[width=1.0\linewidth]{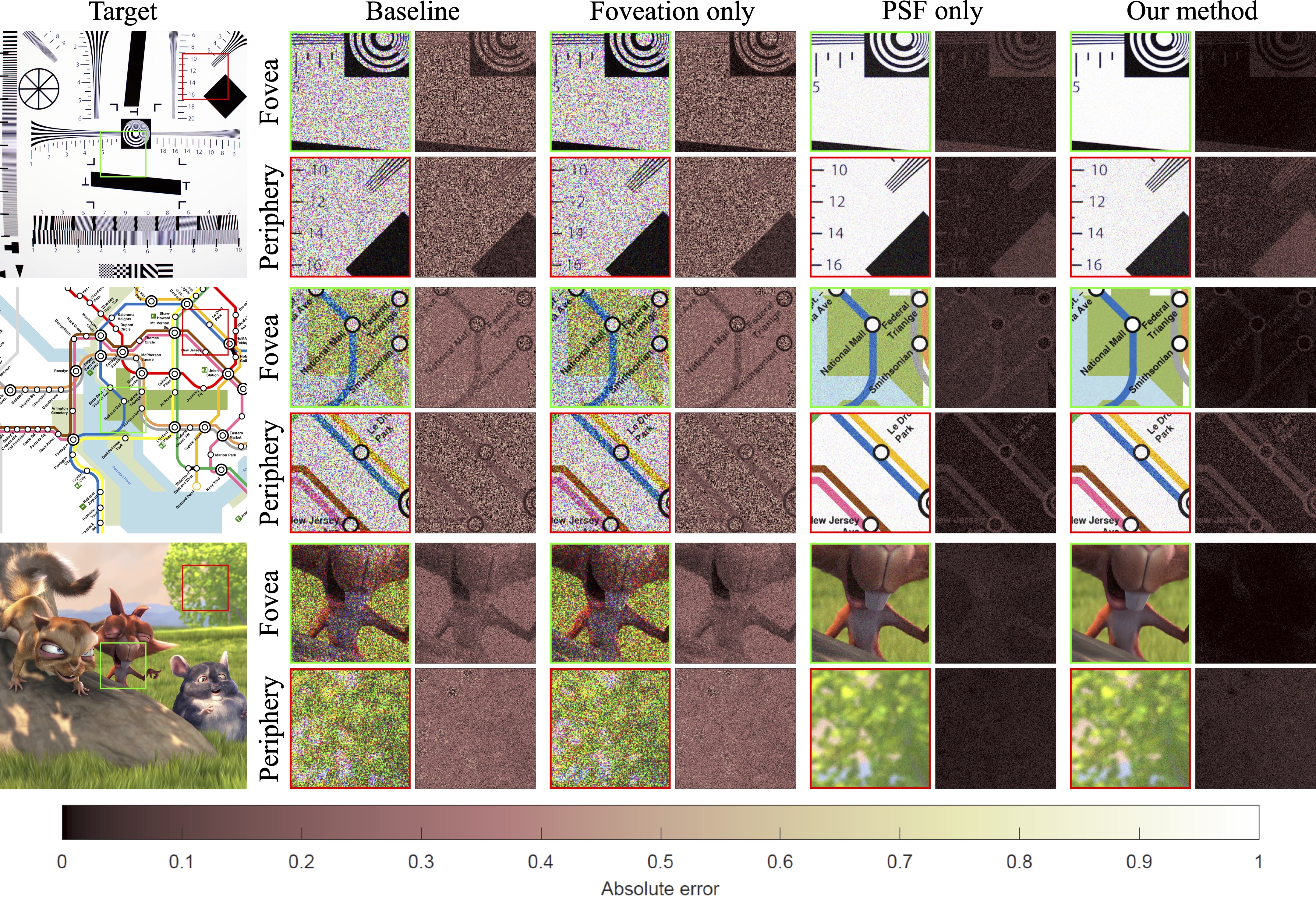}
    \caption{Outputs from different optimization methods for 3 sample input images and the maps of mean absolute error in the reconstructed images. Baseline method assumes uniform visibility of noise across visual field during optimization. Foveation only and PSF only optimizations optimize according to retinal sampling model and PSF of the human eye, respectively. Our model takes into account both the retinal distribution of photoreceptors and PSF in optimization. As a result, we observe smaller reconstruction error due to the use of PSF in optimization and a better distribution of error in fovea and periphery due to placing higher priority on noise reduction in foveal region, where the visibility of the noise is higher.}
    \label{fig:evaluation:condition}
\end{figure*}
In this section, we discuss our approach to reducing perceptible speckle noise in holographic images via anatomically-aware modeling of human perception. Specifically, we introduce the anatomical sampling of retinal ganglion cells and the PSF on retina for modeling the holographic image formation, which we later use for optimizing the SLM phase patterns. We model the RGC and retinal PSF as discussed above, in Section~\ref{sec:hvs_sampling}. Incorporating the retinal PSF into generating holograms helps locally minimize the interference between neighboring pixels, thereby reducing the overall perceptible speckle noise. Additionally, the RGC sampling improves the image fidelity in the fovea, while pushing the noise spatially into imperceptible higher eccentricity peripheral region. 

\paragraph{PSF-induced speckle} 
While the PSF of a every individual human eye needs to be measured and is difficult to model, we simplify it to be a Gaussian~\cite{coppens2004new}. We approximate the superposition of complex amplitude of the neighboring pixels on the retina as the convolution of the per-pixel complex wavefield on the image plane with the Gaussian PSF kernel, as follows

\begin{equation}
    \label{eq:psf}
    PSF(z) \triangleq z*G_{psf},
\end{equation}
where $G_{psf}$ is the Gaussian PSF kernel. For a Fourier holographic projection, the size of the PSF kernel can be approximated using the distance of holographic projection and the angle subtended by the Airy disk, as discussed in Section~\ref{sec:cgh}. 
For our experiments, we use a Gaussian kernel with a standard deviation of $0.6$ for modeling the PSF, as obtained by minimizing the variance between the parameterized Gaussian kernel and the dataset of retinal PSFs.

\paragraph{mRGC guided foveation}

As discussed in \Cref{sec:hvs_sampling}, the density of the mRGC decreases with larger eccentricity. This means that achieving high-fidelity at smaller eccentricities and a compromised image quality at larger eccentricities is sufficient for high perceptual quality, as opposed to eliminating the noise completely from the entire image. To this end, we introduce the analytical and anatomical model \Cref{eq:density} to our phase retrieval method. The goal here is to adaptively distribute the perceived (speckle) noise to each pixel based on the likelihood to be sampled by an mRGC, as follows
\begin{equation}
mask_{foveation}(x, y) \triangleq r/\sqrt{\frac{2}{\sqrt{3}}\left(\frac{x^2}{\rho(r,1)}+\frac{y^2}{\rho(r,2)}\right)},
\end{equation}
where $r=\sqrt{x^2+y^2}$ represents the retinal eccentricity of the point $(x, y)$ in terms of visual degrees. The right side of the equation was discovered by Watson et al. \cite{watson2014formula} to represent the general receptive cell density with only 1st and 2nd types of meridians.

\paragraph{Optimization loss}
Eq.~\ref{eq:basic-optim-problem} describes the basic optimization goal for a standard phase retrieval problem. Our aim here is to develop a unified novel loss model that considers both PSF and foveation as discussed above.
We define our perceptually-aware and PSF-considered penalty function as

\begin{equation}
\mathcal{L}_{ours} = \sum_x \sum_y mask_{foveation}(x, y)|| PSF(z)_{x, y} - I_{x, y} || ^2.
\end{equation}
Here $PSF(z)_{x, y}$ and $I_{x, y}$ represents the intensity of the the reconstructed image and target image at a given $(x, y)$ respectively, on retina.
We replace the penalty $\mathcal{L}$ in Eq. \ref{eq:basic-optim-problem} with our custom loss function to solve the Wirtinger phase retrieval optimization. Note that a more complicated penalty function can also be utilized for perceptually-aware optimization~\cite{chakravarthula2019wirtinger}.

\section{Evaluation}

We evaluate our methods with both image-space objective metrics and perceptually-aware subjective studies. Reconstructed holographic images with our method are compared with state-of-the-art first-order gradient descent optimized holograms~\cite{chakravarthula2019wirtinger,peng2020neural} (BASELINE). Moreover, we also compare with only mRGC guided optimization without considering PSF-induced speckle (FOVEATION-ONLY), and a PSF-only optimization without mRGC foveation (PSF-ONLY). The conditions are described in Table~\ref{table:exp-conditions} and the corresponding results are also visualized in \Cref{fig:evaluation:condition}. 

\begin{figure*}[ht!]
\centering
    \includegraphics[width=0.98\linewidth]{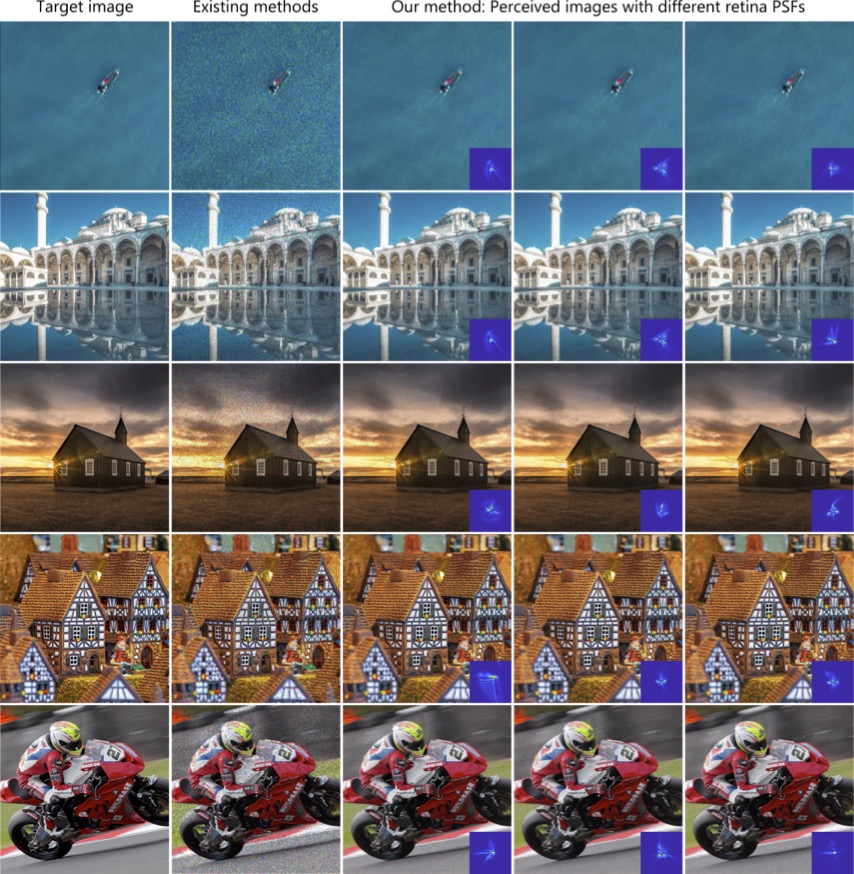}
    \caption{Holograms optimized using existing methods~\cite{chakravarthula2019wirtinger,peng2020neural} do not consider the speckle noise caused by retinal PSF and hence the perceived images are noisy. However, optimizing for individual PSFs which might differ from person-to-person and even from left to right eye is challenging. Here we validate that optimizing hologram phase with a Gaussian PSF consistently suppresses retinal speckle noise and is a reasonable generalization.}
    \label{fig:gauss_opt_retina_recon}
\end{figure*}

Apart from PSF-induced speckle, there are a multitude of reasons for noise and poor image quality in real holographic display systems, such as SLM phase nonlinearities, fringe fields and non-smooth pixel surfaces. Since it is difficult to study the effect of PSF-induced speckle alone, we carry our perceptual experiments only in simulation. As shown in \Cref{fig:noiseVis}, our proposed method suppresses perceptible speckle noise for a variety of retinal PSFs (see \Cref{fig:retina_psf}). However, since every user has a unique retinal PSF, it is challenging to model holograms specific to individual users. Therefore, we instead use a Gaussian PSF as a population average of different real-world retinal PSFs~\cite{coppens2004new}. A standard deviation of 0.6 was obtained by minimizing the variance between the parameterized Gaussian kernel and the retinal psf dataset. It is important to remember though, that coma, astigmatism and other higher-order PSFs cannot be completely approximated using a Gaussian. To validate the Gaussian approximation, we first optimize holograms using a Gaussian PSF and then synthetically reconstruct the perceived retinal images using different real retinal PSF measurements. As shown in \Cref{fig:gauss_opt_retina_recon}, we notice a significant reduction in the perceived speckle noise for a variety of PSF profiles. Table.~\ref{table:psnr_metrics} shows quantitative assessment of the reconstructed holographic images. 

\begin{table}[h]
\centering
\caption{Conditions describing the different methods against which we compare our method. We incorporate both PSF-induced speckle and mRGC guided foveation into the hologram generation method to produce perceptually high quality reconstructions.}
\label{table:exp-conditions}
\begin{tabular}{r|c|c}
\multicolumn{1}{l|}{} & PSF-induced speckle         & mRGC guided foveation                              \\ \hline
Baseline              & \cellcolor[HTML]{FFCCC9}No  & \cellcolor[HTML]{FFCCC9}No                         \\ \hline
Foveation-only        & \cellcolor[HTML]{FFCCC9}No  & \cellcolor[HTML]{9AFF99}Yes                        \\ \hline
PSF-only              & \cellcolor[HTML]{9AFF99}Yes & \cellcolor[HTML]{FFCCC9}No                         \\ \hline
Ours                  & \cellcolor[HTML]{9AFF99}Yes & \cellcolor[HTML]{9AFF99}{\color[HTML]{000000} Yes} \\ \hline
\end{tabular}
\end{table}

\subsection{Simulated perceptual noise analysis}
\label{sec:analysis}
With 6 target images as the references, as shown in study program file (Ref~\cite{userStudyCode}), we compute the loss of mean-squared error (IMMSE) for image-wise noise level, together with perception-based image quality evaluator (PIQE) \cite{venkatanath2015blind} and HDR-VDP \cite{mantiuk05PredVisDiff} for perceptual quality. 
For a uniform \emph{lower-means-better} error metric across all conditions, we inverted the HDR-VDP metric scores, whose otherwise original value is positively correlated to the perceptual image quality. For a fair comparison among different image conditions, we normalized the error with the BASELINE condition (the dashed horizontal line). 

The results are visualized in \Cref{fig:evaluation:metric}. Among all metrics, both OURS and PSF-ONLY conditions demonstrated significantly lower error than BASELINE: both are $95\%+$ lower than BASELINE/FOVEATION-ONLY with IMMSE. The trend is consistent with perceptual metrics: OURS showed $22.9\%$/$65.4\%$ lower PIQE/HDR-VDP error. The FOVEATION-ONLY condition, however, did not show a significant difference from BASELINE. Meanwhile, OURS and PSF-ONLY are noticeably similar with all three metrics (max $8.8\%$ difference). 
We note that quantitative metrics show slightly higher error for OURS compared to the PSF-only condition as the metrics are evaluated over all image pixels evenly. However, we argue that standard image metrics are not the best way of evaluating our technique as they do not fully capture the human visual system properties in the peripheral vision.
Evaluating foveal and peripheral regions separately is challenging due to continuous visual acuity fall-off. Precisely evaluating the perceived quality and noise level with a novel vision-aware metric is an exciting future work while orthogonal to the scope of the current work. 
An alternative and a more direct evaluation method are perceptual experiments. Therefore, we conducted perceptual evaluation which is more important for perceived display image quality.

\subsection{User Study}
\label{sec:study}

To study the perceptual image quality of the end-user, we performed a subjective evaluation with all conditions.
We used remote experimental studies following the safety protocols due to the restrictions imposed by COVID-19. The programmed user studies were sent to the users in a zip package and written instructions were provided to the users for setting up the experiment. One of the authors also monitored the remote experiment via video conference.
The code used for our user evaluations is attached as the Supplementary material with this manuscript.

\paragraph{Task and Stimuli}
Our perceptual study program required the users to input the size and resolution of their display monitor, and the program automatically computed the eye-to-display viewing distance to be maintained, as well as scaled the stimuli images in order to maintain a fixed field of view for all the users. Maintaining the appropriate viewing distance, the users kept their head fixed and one of their eyes covered. The stimuli images were placed at the center of the display. A green cross was shown throughout to help with gaze fixation. For each trial, a reference image was shown followed by pairs of stimuli (temporal interleaving).
Specifically, we used $6$ ($1080\times1080$) different target images and their corresponding reconstructed images from the four test conditions: OURS, FOVEATION-ONLY, BASELINE, and PSF-ONLY. 
The stimuli were shown as images on screen while the subjects were instructed to remain seated and gaze at the image center during the whole study. 

The task was a two-alternative-choice (2AFC). 
Before the experiment of each target image, the subjects were shown ground truth original images for 5 seconds and were asked to examine it as the quality reference. A pair of images were then presented for 1 second each as generated by any two of the four test conditions described. The order of the images presented was randomized. 
The 2AFC task for each trial was to choose the image with less perceived noise compared to the reference target which was then followed by the next trial.
The subjects used the keyboard to record their responses.
Each experiment consisted of $36$ trials, $6$ per target image. Twelve users (mean age $24.1$, $3$ females) participated in the study. All had a normal or corrected-to-normal vision. None of the subjects were aware of the number of conditions, our technique, and the goal of the study. The experiment was conducted with pre-computed test images distributed to users to adhere to the ongoing COVID-19 restrictions.

\begin{figure}[bht]
    \centering
    \subfloat[objective metrics]{\includegraphics[width=0.96\linewidth]{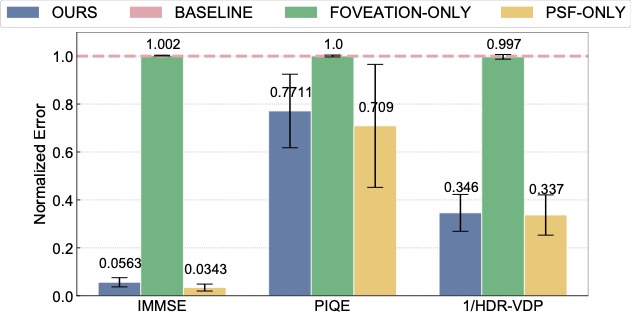}\label{fig:evaluation:metric}}
    
    \subfloat[user study]{\includegraphics[width=0.85\linewidth]{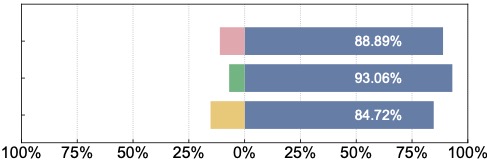}\label{fig:evaluation:study}}
    \caption{Evaluation results of our method. (a) shows the objective errors described in  \protect\Cref{sec:analysis}. Lower error values mean better quality. (b) plots the users' 2AFC choice percentage as being ``less noisy'' from the study in \protect\Cref{sec:study}.}
    \label{fig:evaluation}
\end{figure}

\paragraph{Results}
\Cref{fig:evaluation:study} shows the study results. Comparing with all other conditions, the majority of users chose OURS as the one with less noise. By observing OURS vs PSF-ONLY between \Cref{sec:analysis} and this study, we discovered the significant difference between objective metrics and higher-order end-user perception. That is, the foveation effect in OURS introduces strong perceptual quality gain despite the subtle numerical differences. The result demonstrates the perceptual benefits of both foveation (by comparing with PSF-ONLY) and the introduction of PSF (by comparing with FOVEATION-ONLY) in the optimization.

\paragraph{Ensuring Gaze Fixation without Eye Tracker}
In principle, the best evaluation of our method requires an eye tracker to account for dynamic gaze changes that occur naturally~\cite{lu2020improved}. However, we argue that even without it, our evaluation is valid. This is because the stimuli images have noise in the periphery, and any deviation of the participants’ gaze from the assumed fovea location (green cross) would indeed be disadvantageous for our technique. In practice, such a situation would lead to lower metric scores of our method due to visible noise in regions which were assumed to lie in the periphery. The higher subjective scores for our method, as described below, only confirm that the users maintained gaze fixation and the experiments were conducted appropriately.

\begin{table}[htbp]
	\centering 
	\caption{Quantitative metrics computed on a custom dataset with randomly picked images. It can be observed that optimizing for a Gaussian PSF but reconstructing with a real retinal PSF produces results on par with optimizing directly for the real retinal PSF.} 
	\label{table:psnr_metrics} 
	\begin{tabular}{c|c|c}  
		& & \\[-6pt]  
		 & PSNR & SSIM \\ 
		\hline
		& & \\[-6pt]  
		Baseline & 22.28 & 0.71 \\ 
		\hline
		& & \\[-6pt]  
		Direct retinal optimization & 35.55 & 0.95 \\
		\hline
		& & \\[-6pt]  
		Gaussian PSF optimization & 31.6 & 0.91 \\
		\hline
	\end{tabular}
\end{table}

\section{Discussion and Conclusion}
Perceptually aware techniques for virtual and augmented reality (VR/AR) displays have gained popularity in recent times, both in terms of designing new display optics~\cite{akcsit2019manufacturing} and rendering images on conventional displays~\cite{tursun2019luminance}. While such displays aim at targeting resources to enhance the foveal image quality at the expense of reduced spatial resolution in the periphery, holographic displays uniquely suffer from the perceived noise in the imagery (arising from both speckle and other sources). 

In this work, we have demonstrated an approach to reduce perceived speckle in holographic displays, by incorporating anatomically-informed model of human visual system and perception. Our simulated retinal image reconstructions show that we can reduce the effect of perceivable speckle noise due to neighboring pixel overlap by a significant amount, by incorporating the speckle formation into hologram generation methods. We have further validated our method by comparing several different perceptual metrics. In addition, our subjective user evaluations also demonstrated an improved image quality using our method and a reduced perceptible speckle noise. 

Our method is currently implemented using unoptimized low-level code that prohibits real time performance, due to the iterative phase refinement approach. However, further optimization of code and dedicated hardware for computation, or a machine learning based hologram generation~\cite{eybposh2020deepcgh}, can promise toward real time performance. Due to the ongoing COVID-19 pandemic, our subjective user evaluations were also conducted remotely, for which the holographic reconstructions were pre-computed. In order to generalize for various users, we employed a Gaussian PSF model. However, we note that different imaging systems, including the human eyes, exhibit different PSF properties and incorporating a well characterized PSF of the measurement device (i.e., an eye or a camera) in hologram generation may further improve the perceptual quality of images. 

Our current model considers the retinal sampling density that relates to noise detection than discrimination~\cite{thibos1987retinal,hirsch1989spatial}. Extending our approach to also reflect the spatio-temporal effects of human perception is an exciting future research direction. Recent body of research also suggest that the chromatic aberrations in the eye and the defocus blur, which relates to the PSF on the retina, play a significant role in driving the eye's accommodation~\cite{cholewiak2017chromablur}. Since the eyes can only focus at a single depth at any given time, employing a more generalized definition of PSF (or a well customized PSF per individual user) can potentially help generate 2D holographic projections with realistic defocus cues, thereby guiding accommodation in holographic near-eye displays and eliminating the need to compute true 3D holograms. Moreover, the PSF on the retina can be \emph{holographically} engineered, triggering accommodation responses that might be useful in specific scenarios. For example, a future smart AR display can alert and trigger the accommodation response of a car driver to focus on a nearby obstacle. We believe that there are several possibilities at the intersection of human perception and digital holographic displays. We are excited that our work would inspire future investigation in this intersection.

\acknowledgments{
This work is supported by the European Research Council (ERC) under the European Union’s Horizon 2020 research and innovation program (grant agreement N$^{\circ}$ 804226 – PERDY), NSF grants 1840131 and 1405847, and a generous gift from Intel.}

\bibliographystyle{abbrv}

\bibliography{references}
\end{document}